# An indirect correlation of dielectric properties using optical trapping and dielectric resonance in two different frequency regimes

*Somaditya Sen*

*Physics Department, Indian Institute of Technology Indore, Khandwa Road, MP, India 453552*

Abstract: Dielectric permittivity, $\varepsilon_r$, of materials are often limited to a sub-GHz range using normal LCR meters. In the GHz range the $\varepsilon_r$ can be measured using Vector Network Analyzers and measurement jigs (waveguides) which are specific for different frequency regimes. Hence, to measure $\varepsilon_r$ for the entire frequency range one needs several components and is an extremely costly experiment. However, for applications such as Dielectric Resonator Antennas one need to know the $\varepsilon_r$ and the dielectric loss to estimate the resonant frequency. An indirect method is proposed in this letter to find the $\varepsilon_r$ from Transversely Misaligned Dual-Fiber Optical Trapping at the optical frequency range to approximately estimate $\varepsilon_r$ and thereby understand the correlation between these two regimes of frequency responses.

*Author Email: sens@iiti.ac.in*

The Transversely Misaligned Dual-Fiber Optical Trap (TMD-FOT) technique uses two optical fibres to create an optical trap. The fibers are intentionally misaligned across their width. This means the light from the fibers does not focus on a single point but creates a trapping field that can affect particles at different positions between them. This setup enables high-precision manipulation and trapping of smaller particles, such as inorganic nanoparticles. Various forces, like the optical gradient and scattering forces, act on the trapped nanoparticle, influencing its motion. An asymmetric trap or a twisted angular momentum of the light can cause rotational or orbital motion in the trapped particle. This motion usually results from the optical beam's angular momentum [1][3], which is called optical torque.

Techniques such as optical tweezers interferometry, video microscopy, or polarization-sensitive detection can measure the rotational dynamics or orbital rotation of the trapped nanoparticle. High-precision imaging can reveal the orientations of these particles, helping to determine how they rotate around their centers. By changing the polarization of light or applying an external force field, the movements can be tracked. This allows for applying angular velocity and torque to study friction, interactions with the surrounding medium, and their moment of inertia. Particle-particle interactions can also be explored, which is important for biomedical and optical applications. Among the various motion-related studies, analyzing the rotational Brownian motion of nanoparticles in different media is particularly interesting. Another compelling topic in physics involves light-matter interactions, especially concerning optical vortices.

All the processes described depend on how light interacts with small matter particles, creating radiation pressure and a gradient field from the misalignment of the fibers. A structured light field creates a region where particles can be trapped, thanks to a varying intensity gradient or structured polarization. This allows for precise control over particle motion. Particles are drawn toward the area of highest intensity in the gradient field.

Light can carry both spin angular momentum (SAM) and orbital angular momentum (OAM). SAM relates to the polarization of light, such as circular polarization, while OAM is linked to the helical structure of the light's wavefront. SAM is the intrinsic angular momentum of light determined by its polarization; for example, circularly polarized light carries SAM. In contrast, OAM relates to the spatial arrangement of the light beam and is tied to the helical phase fronts of vortex beams. These beams contain a phase singularity along the propagation axis, with the OAM connected to the number of twists in the helix. This concept is vital in TMD-FOT, where the twist can be transferred to a trapped particle. Helical polarized light, whether circular or elliptical, or a phase gradient, can apply torque to the trapped particle, causing it to rotate around its axis or orbit in a circular path due to the OAM.

It is important to note that the torque relies on the intensity distribution of the light field, which is proportional to the gradient of light intensity, as well as the light's polarization and the particle's material properties. Additionally, another force acts on the particles. Due to the light, the scattering force pushes the particles along the direction of light propagation. The equilibrium position occurs where the scattering force balances the gradient force, resulting in particle trapping. However, extra forces arise from the medium's nature. Stability is typically reached when the forces from the light field counterbalance the forces from the particle's interactions with the surrounding medium, such as Brownian motion. If a particle tries to leave the trap center, the light field pulls it back toward the center. Meanwhile, the optical vortex creates rotational motion, leading to either rotation or orbital motion.

Understanding the principles of light-matter interaction is crucial to achieving efficient nanoparticle trapping conditions. Oxides are among the most common materials in nature, and scientific reports often focus on important oxide systems like ZnO, $TiO_2$, FeO, $Fe_2O_3$, $Fe_3O_4$, CuO, NiO, and $SiO_2$. However, due to their low refractive indices, trapping nanoparticles made from these materials is challenging. Nonetheless, careful misalignment of the optical fibers can enhance the gradient force, allowing for successful trapping. This setup enables the study of these nanoparticles' material properties [5], which vary based on their size, shape, and refractive index.

The trap's stiffness and position can be finely adjusted by changing the misalignment angle and the distance between fibers. This offers dynamic control for force spectroscopy or nanoscale assembly. Fiber-based traps can be inserted into confined or hard-to-reach environments, such as microfluidic channels or biological tissues. Therefore, they are helpful for biosensing or drug delivery studies that use oxide nanocarriers like $Fe_3O_4$ or ZnO. This makes them suitable for remote and in vivo applications.

Additionally, TMD-FOT can extract the dielectric properties, such as permittivity and loss tangent, from the trap's stiffness and scattering behavior. This advancement represents a significant benefit for experimental condensed matter physics. TMD-FOT allows for non-contact sorting or filtering of nanoparticles based on their optical response, which relates to their homogeneity, crystallinity, and composition. It enables selective trapping and isolation of monodisperse, phase-pure, or highly anisotropic particles, resulting in more consistent dielectric properties. Achieving such uniformity is often challenging but essential for microwave applications, as it affects resonant frequency, bandwidth, and Q-factor. Thus, this method also aids in impurity analysis and doping.

An important relationship exists between the optical gradient fields of TMD-FOT and microwave field confinement in dielectric resonator cavities. Examining how particles orient, oscillate, or align under optical field gradients can provide insights into polarization dynamics or anisotropy. Therefore, in antennas using ceramic materials like $BaTiO_3$, the analysis of these materials becomes important, leading to multi-mode or polarization-sensitive dielectric resonator antennas. Using TMD-FOT, microscale dielectric particles, such as spherical $BaTiO_3$ microparticles, can be levitated and tested for scattering, resonance, or coupling behavior. This presents a novel approach to studying dielectric resonator antenna resonant behavior at optical frequencies before scaling to the GHz range. This helps assess morphology, size, and polarizability [8][9], offering a non-invasive pre-fabrication tool to estimate a material's polarizability and dielectric permittivity. However, it is essential to remember that this dielectric permittivity is measured in the UV, visible, or IR light ranges, around 100 to 2000 nm. This corresponds to frequencies of approximately $3 \times 10^{15}$ to $1.5 \times 10^{14}$ Hz, not in the GHz range where the dielectric resonator antenna demonstrates resonant behavior. Therefore, actual GHz permittivity values must be measured using microwave techniques rather than optical trapping.

Nevertheless, the intrinsic material attributes influencing these responses overlap, allowing for indirect correlations. TMD-FOT is a pre-screening or complementary technique instead of a substitute for GHz dielectric measurements [7][10]. Moreover, remember that TMD-FOT only examines isolated particles; GHz permittivity depends on collective effects, composite packing, and interfaces.

The polarizability ($\alpha$) can be linked to the optical trapping force ($F_{trap}$) and the electric field intensity gradient, with $F_{trap}$ related to the light intensity gradient. For an ideal dielectric sphere with permittivity $\varepsilon_p$ in a medium with permittivity $\varepsilon_m$, one can derive: $\alpha$ is proportional to $(\varepsilon_p - \varepsilon_m)/(\varepsilon_p + 2\varepsilon_m)$. If we consider air or vacuum as the medium and let the radius of the sphere be R, $\alpha$ can be expressed as $\alpha = 4\pi R^3 [(\varepsilon_p - 1)/(\varepsilon_p + 2)]$. Thus, by measuring $F_{trap}$ and selecting a specific gradient of intensity, one can estimate $\alpha$ and consequently calculate $\varepsilon_p$.

While TMD-FOT can differentiate trapping stiffness, Brownian dynamics, and asymmetries related to size and shape variations or surface defects, microwave permittivity is sensitive to domain wall motion, which is more significant at the MHz to GHz range, as well as grain size effects and defect density. TMD-FOT helps identify nanoparticles with either suppressed or enhanced polar domain behavior, which affects GHz permittivity. In TMD-FOT, the asymmetric trap stiffness or anisotropic motion indicates directional polarizability, implying crystallographic orientation, local strain, and ferroelectric anisotropy.

Hence, this letter is an indicator of a correlation of dielectric properties of especially oxide ceramic materials using two different probes in two different frequency regimes.

# References

[1] Simpson, S. H., & Hanna, S. (2010). Rotation of optically trapped particles in vacuum. *Optics Express*, 18(21), 21825–21834.

[2] Padgett, M., & Bowman, R. (2011). Tweezers with a twist. *Nature Photonics*, 5, 343–348.

[3] Yao, A. M., & Padgett, M. J. (2011). Orbital angular momentum: origins, behavior and applications. *Advances in Optics and Photonics*, 3(2), 161–204.

[4] Dholakia, K., & Zemánek, P. (2010). Colloquium: Gripped by light: Optical binding. *Reviews of Modern Physics*, 82(2), 1767.

[5] MacDonald, M. P., Spalding, G. C., & Dholakia, K. (2003). Microfluidic sorting in an optical lattice. *Nature*, 426, 421–424.

[6] Svoboda, K., & Block, S. M. (1994). Biological applications of optical forces. *Annual Review of Biophysics and Biomolecular Structure*, 23, 247–285.

[7] Sebastian, M. T., & Jantunen, H. (2008). Low loss dielectric materials for LTCC applications: A review. *International Materials Reviews*, 53(2), 57–90.

[8] Kumar, A., Dey, K. K., & Sen, S. (2021). Ferroelectric–magnetodielectric coupling in $BaTiO_3$–$NiFe_2O_4$ composites and their microwave dielectric behavior. *Journal of Alloys and Compounds*, 858, 157733.

[9] Petrov, P. K., & Alford, N. M. (2005). The effect of microstructure on dielectric properties of ferroelectric ceramics. *Journal of the European Ceramic Society*, 25(12), 2827–2831.

[10] Petosa, A., & Ittipiboon, A. (2010). Dielectric resonator antennas: A historical review and the current state. *IEEE Antennas and Propagation Magazine*, 52(5), 91–116.